\documentclass[preprint]{aastex631}

\shorttitle{Polarized Megamasers}
\shortauthors{Gallimore et al.}

\newcommand{\fpol}{\hbox{$f_{pol}$}}
\newcommand{\water}{\hbox{H$_2$O}}
\newcommand{\ngc}{\hbox{NGC~1068}}
\newcommand{\ngcstar}{\hbox{NGC~1068$^*$}}
\newcommand{\Stokes}[1]{\hbox{Stokes $#1$}}

\newcommand{\E}[1]{\hbox{$10^{#1}$}}
\newcommand\x      {\hbox{$\times$}}

\newcommand\Mo     {\hbox{$M_{\sun}$}}
\newcommand\kms    {\hbox{km\,s$^{-1}$}}

\newcommand\mone   {\hbox{$^{-1}$}}

\newcommand\vlsr   {\hbox{$\mbox{V(LSRK)}$}}
\newcommand\hour{\mbox{$^{\mathrm h}$}}%
\newcommand\minute{\mbox{$^{\mathrm m}$}}%
\newcommand\GI{GI23}
\newcommand\snr{S/N}

\begin{document}
\turnoffedit
\turnoffeditone
\title{The Discovery of Polarized Water Vapor Megamaser Emission in a Molecular Accretion Disk}

\correspondingauthor{Jack F. Gallimore}
\email{jgallimo@bucknell.edu}

\author[0000-0002-6972-2760]{Jack F. Gallimore}
\affil{Department of Physics and Astronomy, 
Bucknell University, 
Lewisburg, PA 17837}

\author[0000-0002-3443-2472]{C. M. Violette Impellizzeri}
\affil{Institute ASTRON, Netherlands Institute for Radio Astronomy, NL-7991 PD Dwingeloo, The Netherlands}
\affil{Leiden Observatory, Leiden University, PO Box 9513, 2300 RA, Leiden, The Netherlands}

\author[0009-0002-1706-7693]{Samaneh Aghelpasand}
\affil{Department of Physics, Alzahra University, North Sheikh Bahaee St., Deh-e-Vanak, Tehran, Iran, 19938-91176}

\author[0000-0002-2581-9114]{Feng Gao}
\affil{Max-Planck-Institut f\"ur Radioastronomie, 
Auf dem H\"ugel 69, 53121 Bonn, Germany}

\author[0009-0006-3231-5252]{Virginia Hostetter}
\affil{Department of Physics and Astronomy, 
Bucknell University, 
Lewisburg, PA 17837}

\author[0000-0001-8975-9926]{Boy Lankhaar}
\affil{Leiden Observatory, Leiden University, PO Box 9513, 2300 RA, Leiden, The Netherlands}
\affil{Department of Space, Earth and Environment, Chalmers University of Technology, Onsala Space Observatory, 439 92, Onsala, Sweden}

\begin{abstract}

For the first time in an extragalactic source, we detect linearly polarized \water{} maser emission associated with the molecular accretion disk of NGC~1068. The position angles of the electric polarization vectors are perpendicular to the axes of filamentary structures in the molecular accretion disk. The inferred magnetic field threading the molecular disk must lie within $\sim 35\degr{}$ of the sky plane. The orientation of the magnetic fields relative to the disk plane implies that the maser region is unstable to hydromagnetically powered outflow; we speculate that the maser region may be the source of the larger scale molecular outflow found in ALMA studies. The new VLBI observations also reveal a compact radio continuum source, \ngcstar{}, aligned with the near-systemic maser spots. The molecular accretion disk must be viewed nearly edge-on, and the revised central mass is $M = (16.6 \pm 0.1) \times 10^6$~\Mo{}.
\end{abstract}

\section{Introduction} \label{sec:intro}

Extragalactic \water{} masers, or megamasers, are usually associated with (sub-)pc-scale, edge-on molecular accretion disks, jet-induced shocks, or molecular outflows in narrow-line active galaxies (AGNs) \citep{2005ARA&A..43..625L}. Famously, where they occur in disks, their distances can be determined geometrically through kinematic parallax \citep[e.g.,][]{1999Natur.400..539H,2013ApJ...775...13H,2016ApJ...817..128G}. Disk megamasers commonly show a Keplerian rotation curve, precisely measuring the central compact mass, presumably a supermassive black hole (SMBH) \citep[e.g.,][]{2011ApJ...727...20K,2017ApJ...834...52G}.

In various models for accretion disks, the magnetic field plays an important role, ultimately providing an effective viscosity that drives accretion \citep{1975ApJ...200..187E,1998RvMP...70....1B,2013LRR....16....1A}. Ordered and inclined magnetic fields can accelerate clouds, generate outflow, and remove angular momentum from the accretion disk \citep{1982MNRAS.199..883B,1992ApJ...385..460E}. \cite{2006ApJ...648L.101E} proposed that the resulting molecular outflow might provide the torus scale height requisite for obscuration-based unification schemes for AGNs \citep[see][for reviews of such unification schemes]{1993ARA&A..31..473A,2017NatAs...1..679R}. As such, measuring the strength and orientation of the magnetic fields in accretion disks contributes to our understanding of the accretion process.

Linear and circular polarization of maser emission measure the magnetic field strength and orientation in molecular gas \citep{1973ApJ...179..111G}. Since \water{} is a non-paramagnetic molecule, its Zeeman splitting in mG magnetic fields is insufficient to produce detectable circular polarization, even in very narrow maser lines \citep{1986ApJ...302..750D,2023A&A...678A..17L}. Despite several attempts \citep[e.g.~][]{1998ApJ...508..243H, 2005ApJ...626..104M, 2020A&A...637A..57S}, \water{} megamaser circular polarization observations have only yielded upper limits to the line-of-sight component of the magnetic field, with the most stringent limit being $B_{\mathrm{los}} < 11$~mG towards NGC~3079 \citep{2007ApJ...656..198V}. Linear polarization of masers is more complicated and depends on the saturation degree, anisotropy of the pumping mechanism, and propagation angle relative to the magnetic field axis, $\theta$ \citep[see, e.g., the discussion in][]{1995AJ....109..507D}. As long as the magnetic precession rate ($\sim 1\ \mathrm{s}^{-1}/\mathrm{mG}$ for the \water{} 22 GHz transition) exceeds the rate of stimulated emission, then the electric polarization vector aligns parallel or perpendicular to the projection of the magnetic field on the sky \citep{1981ApJ...243L..75G}:  when $\theta \la 55\degr{}$, the polarization vector aligns with the projected magnetic field direction; otherwise, the polarization is perpendicular to the magnetic field \citep[cf. the discussion in][]{2006A&A...448..597V}. \cite{2024A&A...683A.117L} demonstrated that anisotropically pumped 22~GHz \water{} masers should show 1 to 3\% linear polarization below the saturation limit, i.e., the limit at which stimulated emission exceeds the decay rate; isotropic pumping produces significant linear polarization only above the saturation limit. Unlike circular polarization, the linear polarization fraction of \water{} masers does not directly constrain the strength of the magnetic field.

The Seyfert~2 galaxy \ngc{} is an unusual \water{} megamaser source. The brightest masers are found in a molecular accretion disk surrounding the central engine, radio continuum source S1 \citep{1996ApJ...462..740G,2001ApJ...556..694G}.  To orient the reader, Fig.~\ref{fig:maseroverlay} shows the sky distribution of the disk maser spots relative to radio continuum. The position-velocity diagram, taken along the major axis position angle ($\mbox{PA} = -50\degr{})$, suggests a rotation curve slightly flatter than Keplerian rotation \citep{gg97}.  However, \cite{2023ApJ...951..109G} (\GI{}) demonstrated that the apparently too-flat rotation curve may be an illusion that results from how the water{} masers preferentially sample spiral arms within the disk; the fully three-dimensional model is consistent with Keplerian rotation around a $17\times10^6$~\Mo{} SMBH. 

The maser disk also shows filamentary substructures; these filaments are especially prominent in the brightest masers, the R4 group. Since the maser region is expected to be partially ionized \citep{1994ApJ...436L.127N,1995ApJ...447L..17N}, the filaments suggest organization by an orderly magnetic field threading the disk (\GI{}). The filament lengths are typically 1--2 mas, or roughly 0.1~pc at the distance of \ngc{}\footnote{The distance to \ngc{} is $13.97\pm 2.1$~Mpc \citep{2021MNRAS.501.3621A}. We adopt the scale 1\arcsec = 70~pc, appropriate for a distance of 14.4~Mpc, to maintain consistency with previous work.}. \GI{} estimated that the magnetic field strengths must be $B \ga 2$~mG to remain orderly in the presence of turbulent motions in the molecular gas. This estimate compares well with an independent measurement from infrared polarimetry, $B \ga 4$~mG \citep{2015MNRAS.452.1902L}.


We obtained new observations of the 22~GHz \water{} megamasers of NGC~1068 to search for polarization as a tool to map the magnetic fields in the molecular accretion disk. As a by-product, we obtained sensitive, wideband continuum measurements that provide the currently highest resolution map of the radio continuum source. Below, we discuss the observations and data reduction, the primary findings, and summarize our main conclusions.

\section{Observations and Data Reduction} \label{sec:data}

We observed NGC~1068 with the High Sensitivity Array (HSA), consisting of the ten VLBA telescopes augmented by the Green Bank Telescope (GBT) and the Karl G. Jansky Very Large Array (VLA) acting as a phased telescope. The observations took place on 2021 Nov 14, starting at UTC 03:36 and lasting six hours. The receivers were tuned to four intermediate frequency bands (IFs) centered at 21.776, 21.904, 22.032, \& 22.160~GHz; the three lower frequency IFs cover the continuum well separated from the \water{} line, and the high-frequency IF centers near the water maser transition $\nu_0 = 22.23508$~GHz, redshifted to $\vlsr{} = 1130$~\kms{}. Observations were performed with full polarization, permitting measurements in parallel-hand circular polarization (right-hand RR and left-hand LL) to produce images in \Stokes{I} and $V$ and cross-hand polarization (RL and LR) to produce images in \Stokes{Q} and $U$. 

\subsection{Data Reduction}

Roughly half of the observing time was dedicated to  NGC~1068, with short gaps for observations of calibrator sources. Every five minutes, the calibrator J0238+16 was observed to align the phases of the VLA antennas. The phase reference calibrator J0239$-02$ was observed for three minutes every half hour to monitor instrumental delays and as an astrometric reference. We also observed the bright radio source 3C~454.3 to calibrate bandpass and multi-band delays, 3C~84 as a low-polarization reference to remove instrumental polarization (leakage-term calibration), and 3C~48 as a polarized reference to calibrate the phase difference between the right-hand and left-hand circular polarization signal path. Note that 3C~84 shows weak, $\sim 0.2$\% linear polarization at 22~GHz \citep{2006MNRAS.368.1500T}, which is small compared to the expected $\sim 3$\% instrumental polarization. We viewed linear polarizations $\la 0.2$\% skeptically as they may reflect systematic error by assuming zero polarization for 3C~84. After accounting for pointing overheads and calibrator observations, the total integration time on NGC~1068 was 2.87 hours.

Data reduction broadly followed the approach used for the previous epoch (HSA code BG262J; \GI{}) with a few extra steps to ensure proper polarization calibration. All calibration steps were performed in AIPS \citep{1999ascl.soft11003A} following the procedures recommended by NRAO. One key step was determining multiband delays to ensure that phase calibration solutions could be transferred from the maser IF to the continuum IFs (AIPS task {\tt VLBAMPCL}). Unfortunately, the multiband delay solutions failed specifically for the Mauna Kea (MK) antenna, so all baselines to that telescope were discarded (flagged) in the final calibrated data set.

After initial calibration of amplitudes, rates, and delays, we re-calibrated phase rates based on the channel containing the brightest \water{} maser, which, for this epoch, had an integrated flux density $S_\nu = 345$~mJy at $\vlsr{} = 1369$~\kms{}. The resulting rate corrections were applied specifically to NGC~1068 and the phase reference J0239$-$02. This correction introduces an astrometric offset that displaces J0239$-$02 from its reference position of about 7~mas; this offset is applied in reverse to determine the astrometric positions of the maser spots in the J2000 reference frame. The expected absolute astrometric precision is $\sim 0.5$~mas, but, because they share a common phase reference, the relative astrometry between the maser spot positions and the 22~GHz continuum is precise, limited mainly by signal-to-noise (\GI{}). In this work, all positions are reported or plotted as offsets relative to RA(J2000) = 02\hour{}42\minute{}40\fs70907, Dec(J2000) = $-$00\degr{}00\arcmin{}47\farcs9444. 

Calibrating circular polarization (\Stokes{V}) is challenging as it is calculated from the difference of bright signals in the RR and LL feeds. We adopted a conservative approach and applied an amplitude self-calibration cycle assuming $V = 0$ for the brightest maser feature \citep[the Zero-$V$ Self-calibration technique of][]{1999AJ....118.1942H}. If the brightest maser has an intrinsic circular polarization, this self-calibration imposes artificial circular polarization on other maser spots and the continuum. In the end, we did not find credible detections of circular polarization; we conclude that any intrinsic circular polarization is below the detection level, $|\Stokes{V}| \la 1$~mJy ($1\sigma$) (see the discussion in \GI{}).

\begin{figure}
\includegraphics[width=\textwidth]{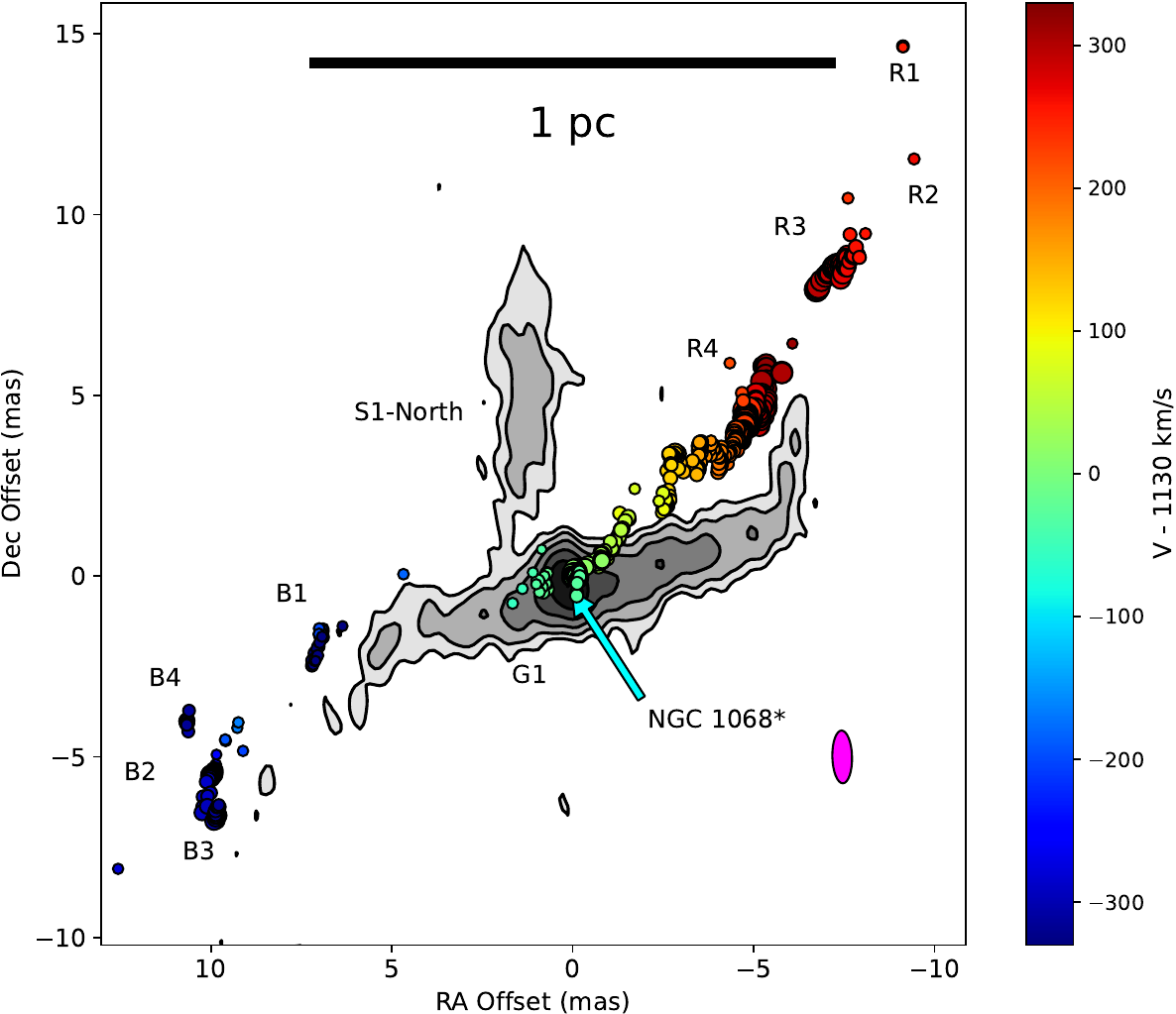}
\caption{An overlay of \water{} maser spot positions and the 21.9~GHz continuum image of the radio component S1 of NGC~1068. Coordinate offsets are relative to RA(J2000) = 02\hour{}42\minute{}40\fs70907, Dec(J2000) = $-$00\degr{}00\arcmin{}47\farcs9444. Maser spots are plotted as filled circles color-coded to the recessional velocity relative to $\vlsr{} = 1130$~\kms{}. The size of the maser spots scales with the square root of the integrated flux density (i.e., larger circles are brighter maser spots). The brightest maser groups (R1, R2, etc.) are labeled using the naming convention of \GI{}. The radio continuum is plotted as filled contours. The contour levels are $-34$ (dashed), $34$, 57, 97, 164, \& 276~$\mu$Jy~beam\mone{}. The equivalent brightness temperatures are $(-0.11, 0.11, 0.18, 0.31, 0.53, 0.88)\times 10^6$~K. The peak surface brightness is $451.5 \pm 0.1\,\mu$Jy~beam\mone{}, equivalent to brightness temperature $(1.45 \pm 0.05)\times 10^6$~K. The locations of the central compact continuum source, \ngcstar{}, and the northern continuum plume, S1-North, are annotated. The continuum restoring beam (FWHM) is the magenta ellipse at lower right.}\label{fig:maseroverlay}
\end{figure}

\subsection{Maser Astrometry and Spectropolarimetry}

After primary calibration in AIPS, all imaging tasks were performed in DIFMAP \citep{2011ascl.soft03001S}. To measure the positions of the maser spots, we used a suitably modified version of {\tt automap}, a script provided with the DIFMAP distribution. The modified script alternates between offset fields associated with radio components S1 and C (the disk and jet masers, respectively) and a control field located midway between the radio components. For maser channels with high signal-to-noise ratios (\snr{}) in \Stokes{I}, we fit the visibilities with Gaussian surface brightness models; we used point source models for channels with lower \snr{}. Imaging was used only to initialize model components and assess residuals; we added model components until the residual peak in the image plane fell below $5\sigma$. To filter false positives, we rejected candidate maser spots whose peak brightness was less than the magnitude of the most negative pixel in the sky plane, and we rejected spots fainter than the brightest pixels in the control field. Finally, we applied a spatial filter to remove false positives associated with sidelobes of the synthetic beam. With natural weighting, the characteristic background rms in the line-free channels is $0.7$~mJy~beam\mone{}, and the synthetic beam size is $1.34 \times 0.52$~mas, PA~$4\fdg2$. For reference, the \href{https://services.jive.eu/evn-calculator/cgi-bin/EVNcalc.pl}{EVN Calculator} predicts an ideal background rms of $0.6$~mJy~beam\mone{}. In the brightest channel, the peak flux density is 280~mJy~beam\mone{}, and the background rms is 1~mJy~beam\mone{}; therefore, the measured dynamic range limit is 280:1. 

\begin{figure}
\centering
\includegraphics[width=0.8\textwidth]{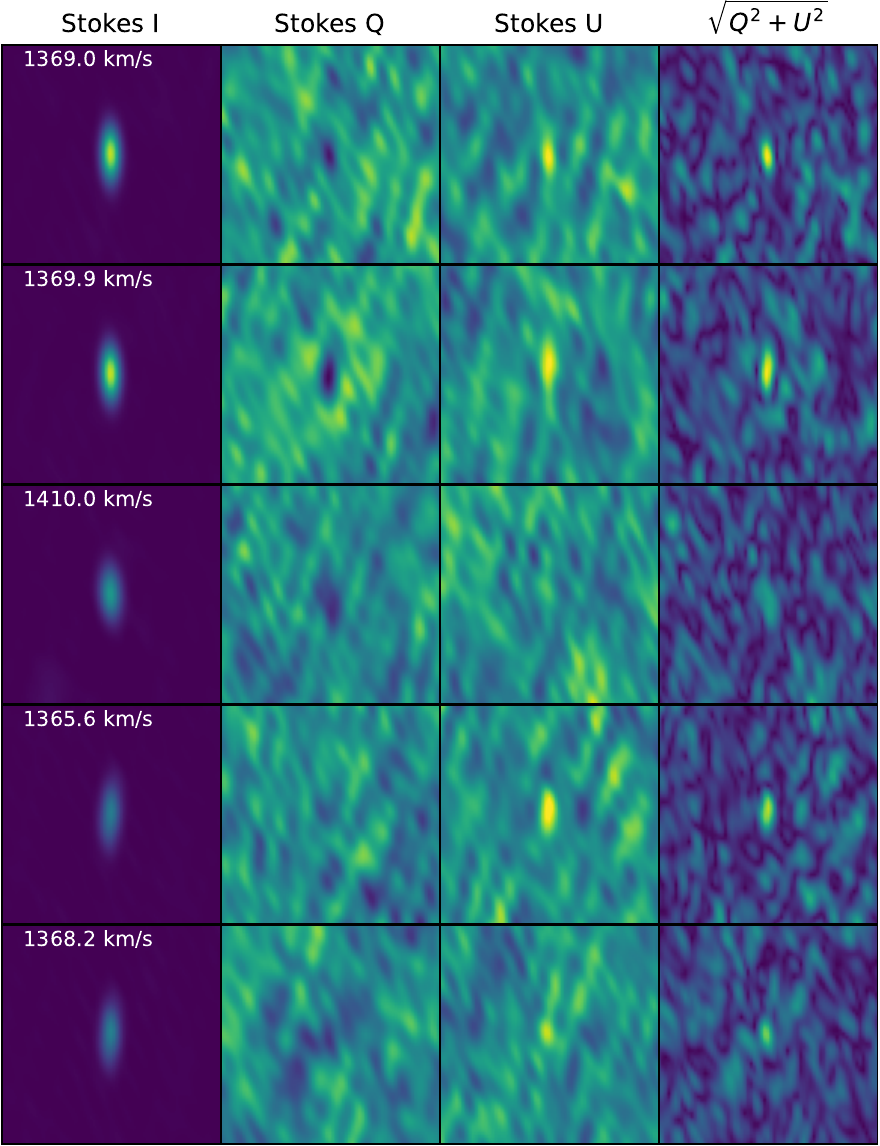}
\caption{Restored and naturally-weighted \Stokes{I}, $Q$, and $U$ images of the five brightest \water{} maser sources in \ngc{}. The rightmost column shows images of the linearly polarized flux density. Each row is a different spectral channel with recessional velocity, \vlsr{}, labeled in the first column. All images are 10~mas square and centered on the brightest maser in that channel. The images are displayed with a positive linear stretch in the viridis (blue-green-yellow-white) colormap. The display stretches are: \Stokes{I}, $(-1,300)$~mJy\,beam\mone{}; \Stokes{Q} and $U$, $(-3,3)$~mJy\,beam\mone{}; and polarized flux, $(0,4)$~mJy\,beam\mone{}. The synthetic beam size (FWHM) is $1.34 \times 0.52$~mas, PA~$4\fdg2$.}\label{fig:polimages}
\end{figure}

\begin{figure}
\centering
\includegraphics[width=0.8\textwidth]{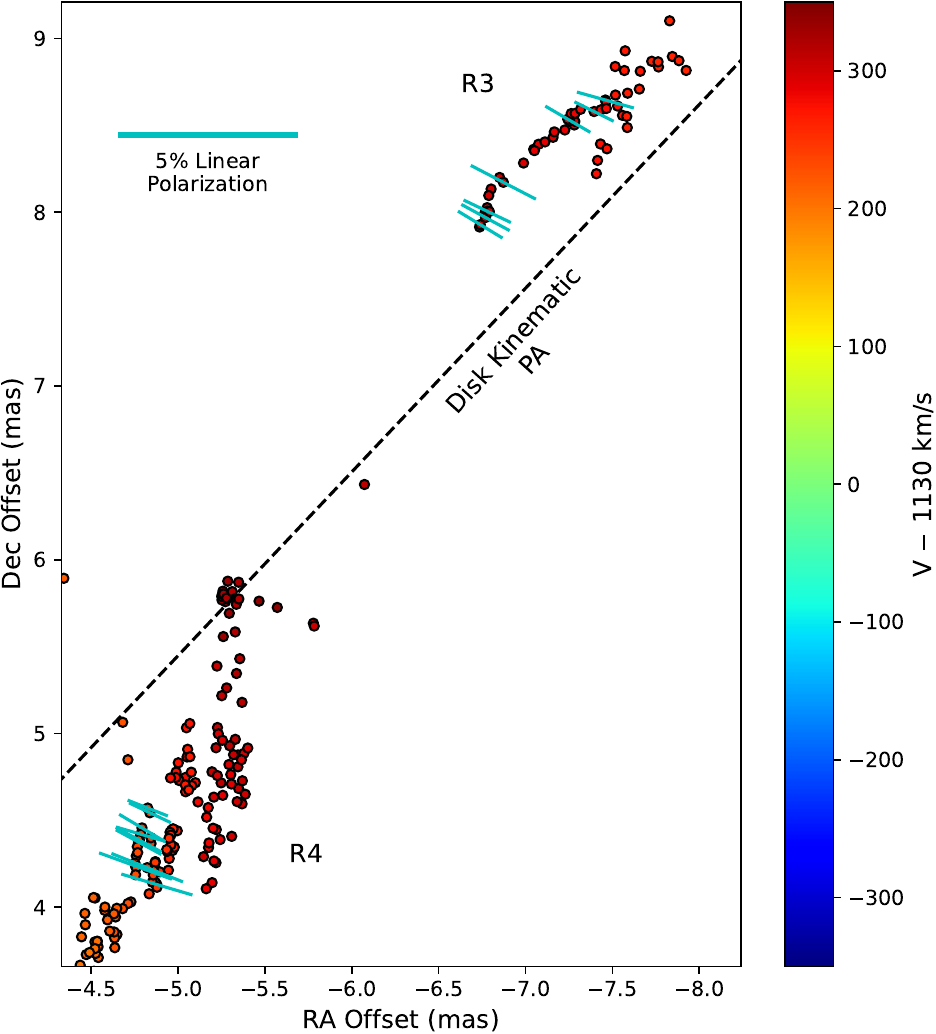}
\caption{Linear polarization map of the \water{} megamasers of NGC~1068. The maser spots are plotted as in Fig.~\ref{fig:maseroverlay} except the spots are now plotted with uniform size. Electric polarization vectors are plotted as cyan lines centered at the position of the peak in the linear polarization maps. The line lengths are proportional to the fractional linear polarization; a $5\%$ linear polarization scale bar is provided at upper left. Polarization vectors are plotted only for those maser spots where the linear polarization fraction is greater than $5\sigma$. The best-fit disk kinematic PA (GI23), an approximation to the major axis of the edge-on disk, is plotted as a black dashed line.}\label{fig:polmap}
\end{figure}

We also generated images in the $Q$, $U$, and $V$ Stokes parameters to search for linear and circular polarization. We used point source models, initialized at the \Stokes{I} positions, to fit the polarized flux. Figure~\ref{fig:polimages} shows the linear polarization images for the brightest five channels in \Stokes{I}. Polarized flux was searched to the limit $\Stokes{I} > 30$~mJy. There were no significant detections on the \Stokes{V} images.  

To assess false detections, we applied a linear regression between the \Stokes{I}, $Q$, and $U$ positions and added a mixture model to identify positional outliers \citep{2010arXiv1008.4686H, 2013PASP..125..306F}. Polarization candidates having an outlier probability $> 10\%$ were rejected; put another way, we rejected polarization candidates whose best-fit position deviated significantly from the \Stokes{I} position. Finally, to obviate the effects of Ricean bias \citep{1974ApJ...194..249W, 1986ApJ...302..306K, 2011JASS...28..267S}, we used a Monte Carlo calculation to evaluate the fractional polarization and the electric polarization vector position angle (EVPA). Fig.~\ref{fig:polmap} shows a map of the maser spots with significant linear polarization, and their properties are summarized in Table~\ref{tab:polarspots}.  All of the polarization detections are associated with the R3d, R4c, and R4d subgroups as defined in \GI{}. After outlier rejection, there are 16 maser spots with fractional linear polarizations $\fpol{} > 5\sigma$ significance. Typical values are $\fpol{} \sim 1.5\%$.

\begin{deluxetable}{cDD@{$\ \pm$}DD@{$\ \pm$}DD@{$\,\pm$}DD@{$\ \pm$}DD@{$\ \pm$}DcD@{$\ \pm$}D}
\tablecaption{\water{} Maser Polarization Data}\label{tab:polarspots}
\tablehead{Channel & 
\multicolumn2c{V(LSRK)} & 
\multicolumn4c{\Stokes{I}} & 
\multicolumn4c{\Stokes{Q}} & 
\multicolumn4c{\Stokes{U}} & 
\multicolumn4c{\fpol{}} & 
\multicolumn4c{EVPA} & 
Subgroup & 
\multicolumn4c{$\theta_{sg}$}   \\
 & 
 \multicolumn2c{(\kms{})} & 
 \multicolumn4c{(mJy)} & 
 \multicolumn4c{(mJy)} & 
 \multicolumn4c{(mJy)} & 
 \multicolumn4c{(\%)} & 
 \multicolumn4c{(degrees)} &
 & 
 \multicolumn4c{(degrees)}}
\decimals
\startdata
597  &  1360.5   &  111 & 1  &  -1.8 & 0.4  &  1.5 & 0.4  &  2.1 & 0.4  &  70 & 6 & R4d & 88 & 6 \\
592  &  1364.8   &  183 & 1  &  -1.4 & 0.4  &  2.1 & 0.4  &  1.4 & 0.2  &  62 & 6 & R4d & 80 & 6  \\
591  &  1365.6   &  198.3 & 0.9  &  -1.5 & 0.4  &  2.3 & 0.4  &  1.4 & 0.2  &  62 & 6 & R4d & 80 & 6 \\
589  &  1367.3   &  188.5 & 0.9  &  -2.2 & 0.4  &  1.4 & 0.4  &  1.4 & 0.2  &  74 & 6 & R4d & 88 & 6 \\
588  &  1368.2   &  191.2 & 0.9  &  -1.3 & 0.4  &  2.4 & 0.4  &  1.4 & 0.2  &  59 & 6 & R4d & 77 & 6 \\
587  &  1369.0   &  345.3 & 0.8  &  -2.9 & 0.4  &  2.7 & 0.4  &  1.2 & 0.1  &  69 & 5 & R4d & 87 & 5 \\
586  &  1369.9   &  345.1 & 0.8  &  -2.7 & 0.4  &  3.3 & 0.4  &  1.2 & 0.1  &  65 & 5 & R4d & 83 & 5 \\
583  &  1372.4   &  118 & 1  &  -1.8 & 0.4  &  1.7 & 0.4  &  2.1 & 0.3  &  68 & 6 & R4c/d\tablenotemark{a} & 85 & 7 \\
573  &  1381.0   &  99 & 1  &  -1.7 & 0.4  &  1.0 & 0.4  &  2.0 & 0.4  &  70 & 11 & R4c & 86 & 11 \\
541  &  1408.3   &  130 & 1  & -1.8 & 0.4  &  1.0 & 0.4  &  1.6 & 0.3  &  70 & 11 & R3d & 74 & 11 \\
539  &  1410.0   &  228 & 1  &  -1.6 & 0.4  &  2.1 & 0.4  &  1.2 & 0.2  &  64 & 6 & R3d & 80 & 6 \\
535  &  1413.4   &  189 & 1  &  -1.3 & 0.4  &  2.3 & 0.4  &  1.4 & 0.2  &  60 & 6 & R3d & 84 & 6 \\
523  &  1423.6   &  105 & 1  &  -1.2 & 0.4  &  1.7 & 0.4  &  2.0 & 0.4  &  63 & 7 & R3d & 81 & 7 \\
521  &  1425.3   &  162 & 1  &  -1.3 & 0.4  &  2.0 & 0.4  &  1.5 & 0.2  &  62 & 6 & R3d & 82 & 6 \\
520  &  1426.2   &  148 & 1  &  -1.3 & 0.4  &  1.6 & 0.4  &  1.4 & 0.3  &  65 & 7 & R3d & 79 & 7 \\
519  &  1427.0   &  171 & 1  &  -1.1 & 0.4  &  2.1 & 0.4  &  1.4 & 0.2  &  59 & 6 & R3d & 85 & 6 \\
\enddata
\tablecomments{Only maser spots with $\fpol{} > 5\sigma$ significance are included in this table. The subgroup is based on cross-identification with maser spots in \GI{}. The EVPA uncertainty includes a 4\degr{} systematic uncertainty added in quadrature to the statistical uncertainty. The position angle difference between the subgroup axis and the EVPA, $\theta_{sg}$, is defined as the minimum of the supplementary angles between the two vectors. }
\tablenotetext{a}{The subgroup identification for this maser spot is ambiguous. We averaged the properties of subgroups R4c and R4d to calculate $\theta_{sg}$.}
\end{deluxetable}

\subsection{Continuum Imaging}

We transferred calibration solutions from the maser IF to generate the continuum image and used natural weighting in the Fourier transform. 
For the image presented in Fig.~\ref{fig:maseroverlay}, we used a sum of elliptical Gaussian and point-source models to reconstruct the continuum image. Fitting was performed interactively in DIFMAP. We added model components one at a time, elliptical Gaussians for brighter, clearly resolved sources, and point sources for fainter, compact sources. We performed non-linear model-fitting using the DIFMAP task {\tt modelfit}. Fifty iterations usually sufficed to minimize $\chi^2$ for a given set of model components.  We stopped adding model components when the sidelobes in the residual map were indistinguishable from background noise, $11\,\mu$Jy~beam\mone{}, which compares well with that predicted by the \href{https://services.jive.eu/evn-calculator/cgi-bin/EVNcalc.pl}{EVN Calculator}, $12\,\mu$Jy~beam\mone{}. Ultimately, the S1 continuum model consisted of six Gaussians and two point sources. The model components were convolved with the restoring beam ($1.46 \times 0.54$~mas, PA~1\fdg{}5) and added to the residual image to create the restored image. For reference, the HSA recovers roughly half of the flux density of S1 compared to a recent 22~GHz VLA measurement: the total HSA flux is $7\pm1$~mJy, compared to $14.7\pm 0.8$~mJy for the VLA \citep{2024MNRAS.52711756M}.

\section{Results and Discussion}

The new continuum image (Fig.~\ref{fig:maseroverlay}) is the currently highest-resolution image of the nuclear radio source. It reveals for the first time a compact radio source at the kinematic center of the molecular accretion disk. Following the convention used for the Galactic Center \citep[Sgr A$^*$,][]{1974ApJ...194..265B} and M~87 \citep[M87$^*$,][]{2019ApJ...875L...1E}, we identify this compact source as \ngcstar{}. There is resolved radio continuum emission extending roughly 6~mas (0.4~pc) east and west of \ngcstar{} along PA $\sim 105\degr{}$. This extended feature broadly resembles the morphology observed in 5 and 8.4~GHz continuum \citep{2004ApJ...613..794G}, except for the sharp bends at the extreme eastern and western ends. As this structure is perpendicular to the local outflow axis, we maintain the interpretation that this extended continuum emission arises from plasma in a hot disk located inside the molecular accretion disk \citep[cf.][]{1997Natur.388..852G,2004ApJ...613..794G}. There is a $\sim 30\degr{}$ rotation in PA between the plasma disk and the maser annulus suggesting a strong warp at $r\sim 0.5$~pc from the central engine.

Labeled S1-North on Fig.~\ref{fig:maseroverlay}, we discovered  a plume of radio continuum emission north of \ngcstar{} that was not resolved at lower frequencies. A hint of this feature is visible in the tapered 22~GHz continuum image of \GI{} (cf. the comparison with the infrared continuum morphology in \citealt{2022Natur.602..403G} \& \GI{}). As it falls along the direction of larger-scale radio jet, we speculate that this feature is a cloud of plasma recently ejected from the central engine. 

A full kinematic model is beyond the scope of this {\em Letter}, but we note that the near-systemic velocity maser spots (the G1 group) closely align with \ngcstar{}. Assuming \ngcstar{} marks the location of the SMBH, the molecular accretion disk is viewed almost exactly edge-on, i.e., inclination $i = 90\degr{}$. Our previous model had no constraints on the kinematic center; rather, we sought to minimize the scale height of the disk, resulting in a best fit $i = 75\degr{}$ (\GI{}). The $\sin{i}$ correction reduces the inferred SMBH mass by 4\%; the revised central mass is now $M = (16.6 \pm 0.1) \times 10^6$~\Mo{}, pending a revised model that includes prior constraints on the location of the kinematic center.

All polarization detections are associated with the brighter masers of the R3d, R4c, and R4d maser subgroups. The fractional polarization is typically between 1 and 2\%. The $5\sigma$ detection limit is about 2~mJy in either \Stokes{Q} or $U$, so non-detections imply $\fpol{} \la 1\%\, (S_I / 200\,\mbox{mJy})^{-1}$, where $S_I$ is the total (\Stokes{I}) flux density. Only 3 / 754 maser spots have a detection threshold $\fpol{} \la 1\%$. Therefore, most of the maser spots might have $\fpol{} \ga 1\%$ but remain undetected in polarized flux owing to limited \snr{}.

From inspection of Fig.~\ref{fig:polmap}, the EVPAs appear nearly perpendicular to the local maser filament axes. To quantify the relative orientation, we cross-identified the maser spot positions and velocities against the subgroup properties found in \GI{}; the subgroup assignments are included in Table~\ref{tab:polarspots}. The best-fit PAs of the R3d, R4c, and R4d filaments are $-15\fdg{}5\pm 0\fdg{}9$, $-18\fdg{}1 \pm 0\fdg{}3$, and $-36\fdg{}5 \pm 0\fdg{}5$, respectively (\GI{}). We define the PA difference $\theta_{sg}$ as the minimum absolute supplementary angle between the EVPA and the PA of the filament; $\theta_{sg} = 0\degr{}$ means the EVPA is parallel to the filament, and $90\degr{}$ means perpendicular. All of the measured values of $\theta_{sg}$ fall within $3\sigma$ of 90\degr{}, and none are consistent with $\theta_{sg} = 0\degr{}$. In other words, the results are consistent with the EVPAs being perpendicular to the filament axes. With the EVPAs of the maser spots consistently aligned perpendicular to the filaments, Faraday rotation here is likely limited to a few degrees at most.

The relative orderliness of the maser spot EVPAs relative to the filaments they trace indicates a common organizing mechanism: the magnetic field. We assume that the filaments trace the projected orientation of the magnetic field lines, much like solar prominences. Under this assumption, the polarization is perpendicular to the magnetic field lines, indicating that the angles between the sight-line and the magnetic field lines are $\theta \ga 55\degr{}$ \citep{2006A&A...448..597V}; in other words, the magnetic field lines lie within 35\degr{} of the sky plane. Additionally, it appears that the polarized filaments are not at right angles with the disk plane as was suggested by the deprojection of the inclined disk models of GI23. Since we now have evidence that the disk is viewed edge-on and that the magnetic fields are oriented near the plane of the sky, projection effects are small, and the apparent sky rotation of the magnetic field relative to the disk plane is likely close to the true, deprojected rotation. In this case, the filaments are rotated by $\la 30\degr{}$ relative to the disk plane (see Fig.~\ref{fig:polmap}). \cite{1982MNRAS.199..883B} demonstrated that a centrifugal outflow results when the poloidal magnetic field falls within 60\degr{} of the disk plane. We speculate that this region might be the source of the larger scale molecular outflow observed with ALMA \citep[e.g.,][]{2014A&A...567A.125G,2016ApJ...829L...7G,2019ApJ...884L..28I,2020ApJ...902...99I}. The filaments are displaced by up to $\sim 1\ \hbox{mas}$ (0.07~pc) from the best-fit disk plane (Fig.~\ref{fig:polmap}), suggesting the masers arise from molecular gas elevated above the edge-on accretion disk. These results are consistent with the proposal of \cite{2006ApJ...648L.101E} that such hydromagnetic disk winds might inflate the scale height of dusty molecular gas around the molecular accretion disk as required by obscuration-based unification schemes. 

\section{Conclusions}

Using the HSA, we have detected linear polarization for the first time in an extragalactic \water{} maser source, the molecular accretion disk of NGC~1068. Furthermore, we have discovered a compact radio source at the kinematic center that we identify as \ngcstar{}. We list below the main results and conclusions.
\begin{enumerate}
\item The near-systemic masers are nearly aligned with \ngcstar{}, indicating that the molecular accretion disk is viewed nearly edge-on. This corrects a previous model and slightly reduces the central mass to $M = (16.6 \pm 0.1) \times 10^6$~\Mo{}.
\item The maser polarization vectors are perpendicular to their respective maser filaments, arguing for a common polarization mechanism. This result is self-consistent with our proposal that magnetic fields organize the maser filaments.
\item The inferred magnetic fields lie within $\sim 35\degr{}$ of the sky plane. Based on the filament morphology, the magnetic fields make an angle of $\la 30\degr{}$ with the plane of the molecular accretion disk.
\item The polarization results are consistent with the requirements for a centrifugally directed, hydromagnetic wind. We speculate that this region of the molecular accretion disk might be the source of the larger scale molecular outflow and potentially elevates molecular gas to create the dust scale height required by obscuration-based unification schemes.
\end{enumerate}

\begin{acknowledgments}
We thank Mark Claussen and the staff at NRAO (Socorro) for help setting up the HSA observations. We are grateful to helpful feedback and suggestions from the anonymous referee. This paper makes use of data obtained from the NSF's VLBA and VLA, operated by the National Radio Astronomy Observatory. The National Radio Astronomy Observatory is a facility of the National Science Foundation operated under cooperative agreement by Associated Universities, Inc. The GBT is part of the Green Bank Observatory, which is a facility of the National Science Foundation operated under cooperative agreement by Associated Universities, Inc. J.F.G. received travel support from the Tressler Fund for Astronomy at Bucknell University.
\end{acknowledgments}

%

\vspace{5mm}
\facilities{HSA, VLA, GBT, VLBA}


\software{AIPS \citep{1999ascl.soft11003A}, astropy \citep{2013A&A...558A..33A,2018AJ....156..123A}, DIFMAP \citep{2011ascl.soft03001S}, emcee \citep{2013PASP..125..306F}}



\bibliography{masers}{}
\bibliographystyle{aasjournal}

\end{document}